\documentclass[article,amsmath,11pt,amssymb,floatfix,superscriptaddress,
showpagenumber, nofootinbib]{revtex4}
\usepackage{graphicx}
\usepackage{epsfig}
\usepackage{bm}
\usepackage{amsfonts}
\usepackage{xcolor}
\usepackage{caption,subcaption}
\input epsf

\begin{document}
\title{\Large Implementation of cosmological bounce inflation with Nojiri Odintsov generalized holographic dark fluid }

\author{Sanghati Saha}
\email{sanghati.saha1504@gmail.com; sanghati.saha2@s.amity.edu}
\affiliation{ Department of Mathematics, Amity University, Kolkata, Major
Arterial Road, Action Area II, Rajarhat, New Town, Kolkata 700135,
India.}

\author{Surajit Chattopadhyay}
\email{schattopadhyay1@kol.amity.edu; surajitchatto@outlook.com}
\affiliation{ Department of Mathematics, Amity University, Kolkata, Major
Arterial Road, Action Area II, Rajarhat, New Town, Kolkata 700135, India.(Communicating author)}
\date{\today}
\begin{abstract}
The current work reports a study on bounce cosmology with a highly generalized holographic dark fluid inspired by S. Nojiri and S. D. Odintsov, 2017, \textit{European Physical Journal C}, \textbf{77}, pp.1-8. The holographic dark fluid that is mostly used for late-time acceleration has been implemented to reconstruct towards realisation of cosmological bounce. We first used the most generalized Nojiri-Odintsov(NO) cutoff to implement the holographic dark fluid. Accordingly, we have reconstructed this dark fluid via some solutions of scale factors. With those solutions, we have explored the evolution of different cosmological parameters. We have examined the effects of each reconstructed parameter in the context of the realization of the cosmic bounce. Next, we use the analytical inferences of the scalar spectral index, tensor-to-scalar ratio, and slow-roll characteristics of the model to study a bounce inflationary scenario. Since inflation is usually associated with the existence of scalar fields, we looked at a possible relationship between NO generalized holographic dark energy and scalar field models. Plotting the evolution of the potential that results from the scalar fields against time. Finally, we investigated the GSL of thermodynamics in the pre-and post-bounce scenarios. \\
\textbf{Keywords:} NO generalized holographic dark energy; Bounce inflation; Cosmological parameters; Bounce realisation.\\
\textbf{AMS classification}: 83F05, 83C56 
\end{abstract}
\maketitle
\section{Introduction}
It is well-discussed in the literature that quantum field considerations may play a fundamental role in the study of the early- and late-universe evolution \cite{Odintsov1}. A unified description of the early-time acceleration and late-time acceleration eras of our Universe can be achieved by using a holographic approach, as discussed in \cite{Odintsov2}. This "holographic unification" is found to correspond with a variety of higher curvature cosmological models, both with and without matter fields. The holographic cut-off, in the formalism of generalized holographic dark energy (HDE), is generalized to depend on where the particle horizon and future horizon and their time derivatives \cite{Odintsov3, Odintsov4, Odintsov5, NO}. 

Gerard 't Hooft first postulated the renowned holographic principle (HP)\cite{holographic principle1}, drawing inspiration from the studies on black hole thermodynamics\cite{holographic principle2,holographic principle3}. The HP asserts that all of the information contained in a volume of space can be represented as a hologram, which correlates to a theory situated on the boundary of that area. This is a contemporary interpretation of "Plato's cave." Leonard Susskind clarified this principle shortly afterward \cite{holographic principle4} using string theory. Additionally, Juan Maldacena\cite{holographic principle5} proposed the well-known AdS/CFT correspondence in 1997, which is the HP's most effective realization. The HP should be the foundation of quantum gravity, according to current widespread opinion. On the other hand, dark energy (DE) has emerged as a major issue in theoretical physics and contemporary cosmology since its discovery in 1998\cite{Supernova1,Supernova2}. The existing evidence point to a dark matter (DM) component that gives rise to galaxies and associated large scale structures (LSS) distributions, and a cosmological constant $\Lambda$ as the origin of DE driving the current period of the accelerated expansion of our universe. The name "$\Lambda$CDM" for this cosmological model refers to the characteristics of its primary constituents. Despite being supported by the observations \cite{DE1,DE2,DE3,DE4,DE5,DE6,DE7,DE8}, the $\Lambda$CDM model has two shortcomings with cosmological constants. Over the past 18 years, many DE models have been presented to address these theoretical challenges; regrettably, the nature of DE is still entirely unknown. The DE problem is thought to be primarily a quantum gravity problem. Consequently, as HP is the foundation of quantum gravity, it may play an essential role to the solution of the DE challenge.
A new DE model known as the holographic dark energy (HDE) model was proposed in 2004 by Miao Li, after the application of HP to the DE problem \cite{Li}.

The longest distance of a quantum field theory is connected to its infrared cutoff, which is associated with the vacuum energy, according to the holographic principle, which has its roots in string theory and black hole thermodynamics\cite{holographic principle1,holographic principle4,holographic principle6}. This factor has been heavily used in cosmological calculations, especially when looking at the late-time era of the universe. These calculations are currently known as holographic dark energy models \cite{DE8,DE9,DE10,DE11,DE12,DE13,DE14,DE15}, and they are also known to agree well with data \cite{DE16,DE17,DE18,DE19,DE20}. At this point, we should note that the holographic dark energy with the Nojiri-Odintsov cut-off\cite{DE21} is the most generic one, and it's noteworthy to mention that it can also be used with covariant theories\cite{DE22}. In addition to the dark energy model, it is discovered that the holographic energy density is useful in realizing the inflationary history and other early Universe evolutions \cite{DE23,DE24,DE25,DE26,DE27}. In the context of the effective field theoretical method, the first investigation into whether Higgs inflation adheres to the holographic principle was conducted in\cite{DE23}. The holographic model has an advantage in the setting of inflation in that the holographic energy density is naturally large to successfully initiate the inflationary era since the maximum distance (or the cut-off of the theory) of the early Universe is tiny. Furthermore, some of our authors extended the application of the holographic principle from the early Universe studies to the bouncing scenario in\cite{DE28}, demonstrating that the holographic energy density violates the null energy condition, a prerequisite for bounce \cite{holographic bounce1,holographic bounce2, holographic bounce3}, which in turn produces the Universe's bouncing behavior. A comprehensive explanation of the bouncing inflationary era and the Nojiri Odintsov generalized holographic dark energy epoch has not yet been attempted, despite the holographic principle having been heavily applied separately during both the early and late times of the Universe's history. Although the holographic scenario has been extensively utilized in late-time cosmology during the dark-energy era to elucidate the universe's acceleration. The particle and future horizons, the age of the universe, the inverse square root of the Ricci curvature, or a combination of the Ricci and Gauss-Bonnet invariants are the options available for the infrared cut-off in the late-time application for an accelerating universe. Each of the aforementioned values and their derivatives can be arbitrarily combined to create a generic infrared cut-off. Moreover, the infrared radius depends on the singularity time if a Big Rip singularity causes the universe's existence to be finite. The matter and holographic dark energy sectors are both included in the late universe scenario, but the matter sector can be disregarded in the early universe scenario. When it comes to explaining the late-time universe, the holographic dark energy theory works well and produces results that are consistent with evidence.

Regarding early-time cosmology, the holographic principle holds true for all of the aforementioned decisions influencing the evolution of the universe. Early studies of the matter bounce model have been conducted by several researchers \cite{holographic bounce4, holographic bounce5}. In this picture, there is no indication of a mathematical singularity as the universe bounces its way from an accelerated collapse to an expanding epoch. The universe might behave cyclically in this manner. Following the bounce, the universe enters into a matter-dominated expansion. In \cite{holographic bounce6, holographic bounce7}, general $F(R)$ and $F(T)$ models of gravity in the presence of bounce cosmology were studied. Since the nonsingular behavior resolves the cosmic singularity problem in a way and offers an alternative to inflation theory, the application of holographic theory concerning a bounce model is of physical relevance \cite{inflation1, inflation2, inflation3, inflation4, inflation5}. In \cite{inflation6,inflation7}, bounce cosmology was examined using an inhomogeneous fluid model, and in \cite{holographic bounce}, a bounce picture resulting from the use of the holographic principle in the early universe was examined. A generalized future event horizon cut-off model is the reason behind the bounce's present cut-off concept.

\textcolor{black}{Cosmology has a long history of introducing viscosity coefficients, although the physical importance of these phenomenological parameters has traditionally been assumed to be weak or at least subdominant. When the temperature reached roughly $10^{10}$ K during the neutrino decoupling (end of the lepton epoch), it is thought that viscosity had the greatest impact on the very early universe. From a particle physics perspective, Misner \cite{viscosity1} introduced viscosity first; see also Zel'dovich and Novikov \cite{viscosity2}. However, the concept of viscosity was initially introduced much earlier from a phenomenological perspective; Eckart's work was the first of this kind \cite{bulk1}. In recent years, viscosity theories have attracted more attention in cosmology for a variety of reasons, probably most notably from a basic perspective. In \cite{bulk2}, the brane-bulk energy exchange term was demonstrated in viscous brane cosmology. They were interested in the viscous fluid on the brane's energy conservation equation. It actually turned out that the emission process equates to a negative entropy change for the thermodynamic subsystem on the brane as compared to the appropriate equation obtained from the Boltzmann equation. The authors of \cite{bulk3} developed a formula for the entropy of a multicomponent correlated fluid, which, in certain circumstances, simplifies to the Cardy-Verlinde form, which establishes a relationship between the entropy of a closed FRW universe and its energy including Casimir energy. There was an inhomogeneous equation of state that the generalized viscous fluid followed.  It is shown that in certain exceptional instances, such an equation reduces to the classic Cardy-Verlinde formula corresponding to the $2D$ CFT entropy. A viscous Little Rip cosmology is here suggested in \cite{bulk4}. Although bulk viscosity generally favors the Big Rip, the authors discover that there are some circumstances in which the formalism easily adapts to the Little Rip scenario. They demonstrated, specifically, that a Little Rip cosmology can be produced as a pure viscosity effect in a viscous fluid, or, equivalently, in one with an inhomogeneous (imperfect) equation of state. A thorough investigation is carried out as well regarding the combined effect of viscosity and a generic (power-like) equation of state. A recently developed viscous Little Rip cosmology in an isotropic fluid is reported in \cite{bulk5}. Their work concentrated on describing the fluid in the late universe in a turbulent state. They looked at how Universe could evolve from a viscous era with constant bulk viscosity to a turbulent era in the one-component case. In \cite{bulk6}, it is examined how an initial fraction $f$ of turbulent kinetic energy in the cosmic fluid affects the cosmological development in the late quintessence/phantom universe through the decay of isotropic turbulence in ordinary hydrodynamics, in FRW universe.} We shall apply the holographic principle to the early-time bounce epoch in this study. Starting with the Nojiri-Odintsov generalized holographic dark fluid as the background fluid and presuming that the fluid has bulk viscosity, we will examine the evolution of the cosmic fluid using the fundamental bounce model described in \cite{holographic bounce}. By taking into account viscous generalized holographic dark fluid, we link infrared cutoffs to future event horizons and create the related versions of the energy conservation formulas. This allows us to demonstrate an equivalency between our model of viscous fluid bounce and the holographic bounce model, which has the Nojiri and Odintsov's unique cutoff option\cite{NO}. Finding a correlation between inflationary scenarios after the Big Bounce is the major goal of this study. The analysis of bounce inflationary dynamics in diverse theoretical frameworks is the focus of this paper. Traditionally, inflation was initially examined within the framework of scalar-tensor theory in its most basic form. Following the acquisition of the model's constraint range for bouncing parameters, we offer comprehensive computations for the slow-roll indices for inflation in the post-bounce scenario. Next, we have discovered the correspondence between canonical scalar field, tachyonic scalar field, and k-essence which correspond to viscous non-interacting generalized holographic dark energy with Nojiri-Odintsov cutoff.  To ensure that the bounce inflationary scenario is complete, we first examine the scalar-to-tensor ratio and the spectral index of primordial curvature perturbations.

Additionally, we look into this specific model's thermodynamic nature and stability in the current context. We specifically investigate the nature of the universe's total entropy when it is encircled by a cosmic event horizon. To be thorough, we expand the non-interacting Nojiri-Odintsov generalized holographic dark energy (NOGHDE) model to include the scenario in which there is no radiation fluid.
The structure of the paper is as follows. A NOGHDE bouncing model with future event scale as the IR cutoff is presented in the Section-II.
Furthermore, in this section the outcomes of taking into account the non-interaction between the universe's dark sectors are examined using every cosmological parameters. Nojiri-Odintsov generalized holographic dark fluid-driven bounce-inflation is determined in Section-III. In Section-IV, we also explore the thermodynamical properties
of the present model. Finally, in Section-V, we summarize the
conclusions of this work.

\section{An Overview of  Nojiri-Odintsov Generalized Holographic Dark Fluid in Bouncing Framework}
To learn concerning the equation of state of holographic dark energy, we will look into \cite{NO}. Additionally, we are going to discuss about a particular model of generalized holographic dark energy called NOGHDE, which has the Nojiri-Odintsov cut-off combination. Holographic dark energy is given by
\begin{equation}
\rho_{HDE}=\frac{3\mathcal{C}^2}{\mathcal{L}^2}
\label{HDE}
\end{equation}
where, $\mathcal{C}$ be the numerical constant and ($\mathcal{L}$) being the infrared cut-off. To formulate NOGHDE, we have considered the combination:
\begin{equation}
\frac{\mathcal{C}}{\mathcal{L}}=\frac{1}{\mathcal{R}_h}[\tau_0 + \tau_1 \mathcal{R}_h + \tau_2 \mathcal{R}_h^2 ].
\label{infrared}
\end{equation}
Therefore,
\begin{equation}
\rho_{NOGHDE}=3 \left(\frac{\tau_0 + \tau_1 \mathcal{R}_h + \tau_2 \mathcal{R}_h^2 }{\mathcal{R}_h}\right)^2
\label{NO}
\end{equation}
Here, $\mathcal{R}_h$ being the future event horizon, defined by
\begin{equation}
\mathcal{R}_h= a \int_{t}^{\infty} \frac{1}{a} \,dt,
\label{future-event}
\end{equation}
where, $a$ denotes the scale factor. \cite{NO} provides a detailed discussion of the reasoning behind taking these cut-off into account, demonstrating how phantom cosmology may enable the unification of the early- and late-time universes. Moreover, one of the intriguing findings (among the others) covered in \cite{NO} has to do with the potential for phantom {non-phantom transition}, which manifests in a way that suggests the universe may have phantom equation of state both early in time and late in time. In general, there could be a number of phantom and non-phantom phases in the oscillating universe. Nevertheless, an additional example of the cut-off has been proposed and considered in \cite{NO}. It is possible to interpret the cut-off form provided by Eq. (\ref{future-event}) as a specific illustration of the general form of $\frac{\mathcal{C}}{\mathcal{L}}=\frac{1}{\mathcal{R}_h}\sum_i \tau_i \mathcal{R}_h^i$. It is easily discernible that $\dot{\mathcal{R}}_h = H \mathcal{R}_h -1$, which will be utilized for attaining future event horizons when the results are presented in the relevant parts. The FRW metric for a flat universe that is homogeneous and isotropic is given by
\begin{equation}
ds^2=-dt^2+a^2(t)(dr^2+r^2(d\theta^2+sin^2\theta d\phi^2),
\label{FRWmetric}
\end{equation}
in which $t$ represents the cosmic time and $a(t)$ is the scale factor. In this case, the units are $8 \pi G = 1$. For isotropic and homogeneous cosmologies, any dissipation process in an FRW cosmology is scalar. One well-known consequence of the FRW cosmic solutions, which correspond to universes with perfect fluid and bulk viscous stresses, is the possibility of DEC violations \cite{dec1,dec2}. According to Eckart's theory \cite{bulk1}, the bulk viscous pressure $\Pi$ can be determined as follows: $\Pi=-3H\xi$, where $\xi$ is the term characterizing the bulk viscosity of the fluid and can be described as a function of the Hubble parameter because of the transport coefficient. In \cite{xi1} and \cite{xi2}, the bulk viscosity coefficient is assumed in terms of the Hubble parameter $H$. Although, In this study, we have checked the bounce realisation with the background fluid NOGHDE. Before bounce $H$ is negative and after the bounce $H$ is positive. As a consequence the bulk viscous pressure will become sign-changeable as we transit from pre to post bounce scenario. To get rid of this, we consider the bulk viscous pressure as  $\Pi=-3 \xi H^2$. Following that, the two Friedmann equations are expressed as
\begin{equation}
    H^2=\frac{1}{3}\rho_{effective}
    \label{1st-Fried}
\end{equation}
and, \begin{equation}
    6 \frac{\ddot{a}}{a}=-(\rho_{effective}+3 P_{effective})
    \label{2nd-Fried}
\end{equation}
where, $\rho_{effective}=\rho_m+\rho_{NOGHDE}$. Here, we have considered, pressure-less dark matter whose density is denoted by $\rho_m$ and $P_{m}=0$. The effective pressure, $P_{effective}$, is just redefined by the bulk viscosity, causing dissipation, according to $P_{effective} = P_{NOGHDE}+\Pi$ , where, $\Pi$ is denoted as bulk viscosity. According to this, $\rho_{effective}=\rho_m+\rho_{NOGHDE}$. The dark matter under consideration is pressure-free, and its density is indicated by $\rho_m$ and $P_{m}=0$. According to $P_{effective} = P_{NOGHDE}+\Pi$, where $\Pi$ represents bulk viscosity, dissipation results from the effective pressure, $P_{NOGHDE}$, be the thermodynamic pressure. \textcolor{black}{In a nutshell, viscous cosmology is dominated by the Friedmann equations, along with fluid’s equation of state. The present work aims to investigate the implications of adding a bulk viscosity to the formalism for many aspects of cosmological theory. We discuss the very early (inflationary) universe in the following section after emphasizing the fundamental formalism in the remaining portion of this section. We describe the traditional inflation theory in brief, addressing post-bounce scenario, and we derive a number of inflationary observables.}

\subsection{\textcolor{black}{Classification of Bouncing Scale Factor}}
In this matter bounce scenario, the initial singularity seen in both the inflationary and standard Big Bang cosmologies may be properly avoided. In the bouncing model, the Universe first experiences a contraction phase that is dominated by matter, followed by a non-singular bounce, before causal production for fluctuation \cite{bounce1,bounce2,bounce3,bounce4}. An appropriate way to escape the initial singularity observed in both inflationary and ordinary Big Bang cosmologies is in this matter bounce scenario. According to the bouncing model, the Universe undergoes a contraction phase composed primarily of matter, a non-singular bounce, before causal production for fluctuation \cite{bounce1,bounce2,bounce3,bounce4}. Furthermore, the scale factor decreases ($a(t) <0$) during the contracting phase and increases ($a(t)> 0$) during the expanding phase. The scale factor finally equals zero during the matter bounce epoch. Thus, in bouncing cosmic theory, by crossing zero ($H = 0$), the Hubble parameter H moves from negative values $H<0$ to positive ones $H>0$ \cite{bounce5,bounce6,bounce7}. The purpose of this article, under the framework of NOGHDE, is to provide the most recent advances in the topic of dark energy, bouncing cosmology, and the inflationary epoch. Here, we have considered a bouncing scale factor from\cite{holographic bounce},
\begin{equation}
a(t)=a_0 ( A + B t^2)^\nu .
\label{scale factor}
\end{equation}
Therefore, Hubble parameter,
\begin{equation}
H=\frac{2 B t \nu }{A+B t^2}.
\label{Hbounce}
\end{equation}

\begin{figure}
\includegraphics[width=.5\linewidth]{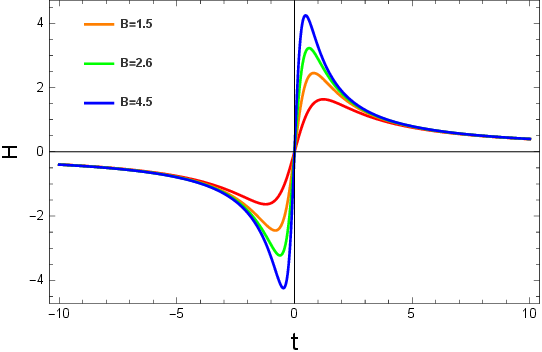}
\caption{Evolution of the bouncing Hubble parameter.}
\label{Fig1_H}
\end{figure}
The integration constants in this case are $A$ and $a_0$. A bounce realization at $t = 0$ is, in fact, described by the scale factor Eq.(\ref{scale factor}). Furthermore, $a(t)$ acts power-law, that is, $a(t) \sim t^{2\nu}$, away from the bounce point, just like a perfect fluid with a uniform EoS parameter. Although, we have fixed the even values of $\nu$ in this whole bouncing case. We have depicted the evolution of the Hubble parameter with regard to cosmic time in FIG.\ref{Fig1_H}. In the pre-bounce scenario, $H$ is negative, indicating a contraction of the universe; at the bouncing point, $H=0$; and in the post-bounce scenario, $H>0$, indicating an expansion of the universe. \textcolor{black}{However, we are assuming bulk viscosity only to the fluid as a whole. Here, it should be noted that the addition of a bulk viscosity is done solely on a phenomenological basis because viscosity depends on the total pressure of the cosmic dark fluid.} 

\subsection{\textcolor{black}{Cosmology of Bouncing Nojiri-Odintsov Generalized Holographic Dark Fluid as Single Fluid}}
In this subsection, we examined at the bouncing non-interacting scenario. In this section, we assume that there is no energy exchange via an interaction term between the two dark fluids of the universe: pressure-less DM and NOGHDE. As a result, the conservation equations' combined form is:
\begin{equation}
\dot{\rho}_m+3H\rho_m=0
\label{conservation_DM}
\end{equation}
\begin{equation}
\dot{\rho}_{NOGHDE}+3H\rho_{NOGHDE}(1+w)=0
\label{conservation_DE}
\end{equation}
where, $w$ be the EoS parameter. In the case of a bouncing universe from Eq.(\ref{scale factor}), we have taken into consideration the density of NOGHDE from Eq.(\ref{NO}), where $\mathcal{C}$ is the model parameter. Again, by using Eq.(\ref{future-event}) to calculate $\dot{\mathcal{R}}_h = H \mathcal{R}_h-1$, and using Eq.(\ref{Hbounce}) to solve it, we can obtain the solution of $\mathcal{R}_h$ and which is:
\begin{equation}
\mathcal{R}_h=\left(A+B t^2\right)^{\nu } C_1-t \left(1+\frac{B t^2}{A}\right)^{\nu } {}_2F_1\left[\frac{1}{2},\nu ,\frac{3}{2},-\frac{B
t^2}{A}\right]
\label{Rh}
\end{equation}
\textcolor{black}{Therefore, we can evaluate the equation of state of NOGHDE in this matter. Thus, from Eq.(\ref{NO}) and Eq.(\ref{Rh}), we can evaluate the energy density of NOGHDE in the following form,
\begin{equation}
\begin{array}{c}
\rho_{NOGHDE}=\frac{3}{\left(C_1 \left(A+B t^2\right)^{\nu }-t \left(1+\frac{B t^2}{A}\right)^{\nu
} {}_2F_1\left[\frac{1}{2},\nu ,\frac{3}{2},-\frac{B t^2}{A}\right]\right)^2} \bigl((\tau _0+(C_1 \left(A+B t^2\right)^{\nu }\\-t \left(1+\frac{B t^2}{A}\right)^{\nu } {}_2F_1\left[\frac{1}{2},\nu
,\frac{3}{2},-\frac{B t^2}{A}\right]) \tau _1+\left(C_1 \left(A+B t^2\right)^{\nu }-t \left(1+\frac{B t^2}{A}\right)^{\nu } {}_2F_1\left[\frac{1}{2},\nu
,\frac{3}{2},-\frac{B t^2}{A}\right]\right)^2 \tau _2){}^2\bigr).
\end{array}
\label{density NOGHDE}
\end{equation}}
Thus, we may derive the density of dark matter pressure-less using Eq. (\ref{conservation_DM}),
\begin{equation}
\rho_m=\left(A+B t^2\right)^{-3 \nu }\rho_{m_0}.
\label{density darkmatter}
\end{equation}

\begin{figure}
\centering
\begin{minipage}{.5\textwidth}
  \centering
\includegraphics[width=.7\linewidth]{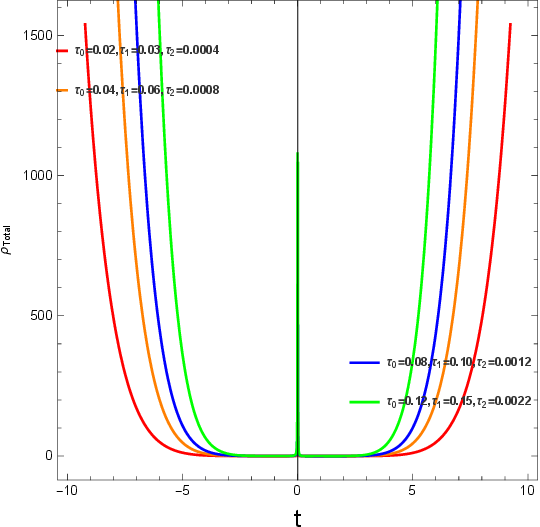}
\caption{Evolution of reconstructed Nojiri Odintsov generalized dark fluid density in viscous non-interacting bouncing scenario.}
\label{Fig2_rho}
\end{minipage}%
\begin{minipage}{.5\textwidth}
\centering
\includegraphics[width=.7\linewidth]{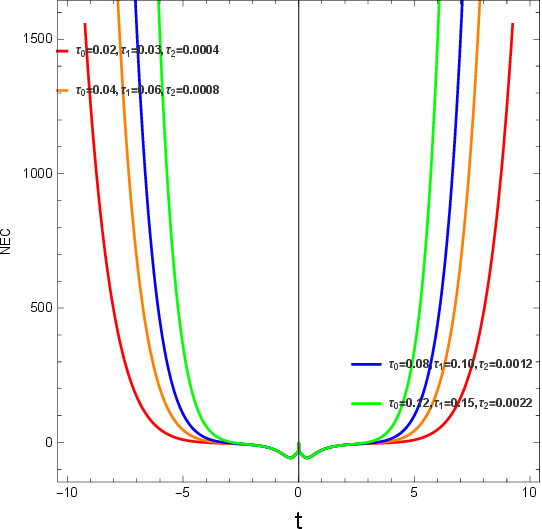}
\caption {Evolution of NEC for the reconstructed background Nojiri Odintsov generalized dark fluid in viscous non-interacting bouncing scenario.}
\label{Fig3_NEC}
\end{minipage}
\centering
\begin{minipage}{.5\textwidth}
  \centering
\includegraphics[width=.7\linewidth]{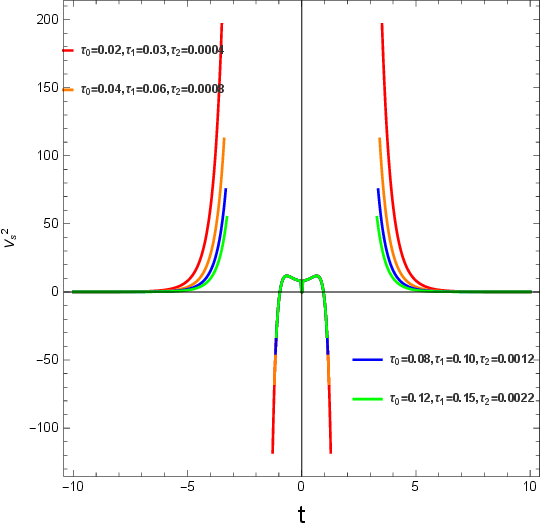}
\caption{Plot of square-speed of sound test $(V_s^2=\frac{dP_{effective}}{{d\rho_{effective}}})$ in viscous non-interacting bouncing scenario.}
\label{Fig4_Vsquare}
\end{minipage}%
\end{figure}

\begin{figure}
\centering
\begin{minipage}{.5\textwidth}
  \centering
\includegraphics[width=.9\linewidth]{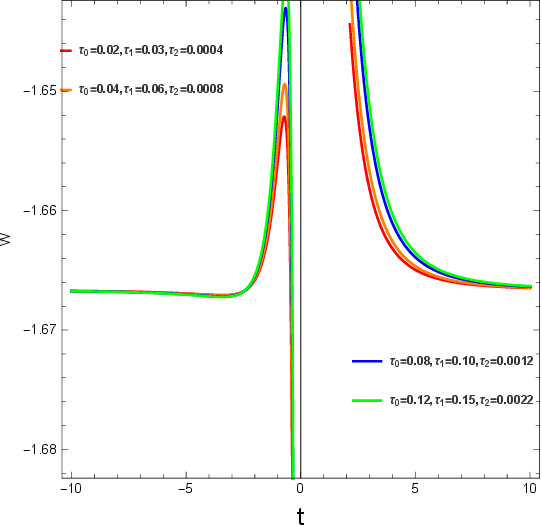}
\caption{Evolution of reconstructed EoS parameter with respect to time in bouncing viscous non-interacting NOGHDE scenario.}
\label{Fig5_W}
\end{minipage}%
\begin{minipage}{.5\textwidth}
\centering
\includegraphics[width=.9\linewidth]{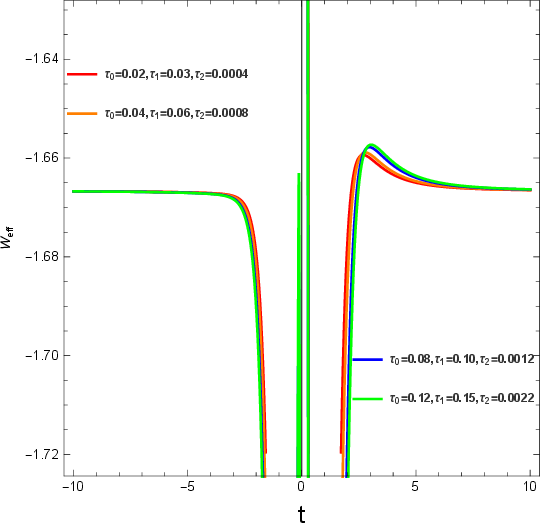}
\caption {Evolution of reconstructed viscous effective EoS parameter with respect to time in bouncing viscous non-interacting NOGHDE scenario.}
\label{Fig6_Weff}
\end{minipage}
\end{figure}

Thus, we can obtain the effective density of non-interacting NOGHDE fluid by $\rho_{effective}$ from Eqs. (\ref{density darkmatter}) and (\ref{density NOGHDE}) and plotted in Fig.\ref{Fig2_rho}(Left Panel),
\begin{equation}
\rho_{effective}=\rho_m+\rho_{NOGHDE}
\label{total density}
\end{equation}
Again, we can write the conservation equation in the form of
\begin{equation}
\dot{\rho}_{effective}+3H\rho_{effective}(1+w_{effective})=0,
\label{conservation_effective}
\end{equation}
where, $\rho_{effective}=\rho_m+\rho_{NOGHDE}$ and $P_{effective} = P_{NOGHDE}+\Pi$. 
Although from Eq.(\ref{2nd-Fried}), we can get the expression of total effective pressure of induced by non-interacting viscous NOGHDE dark fluid. The Null Energy Condition\cite{nullenergy} states that energy density cannot be negative as gravity is always attractive. It is represented by a series of linear equations that include the total effective pressure value and total background fluid density. NEC can never be negative. To achieve a successful non-singular bounce and, hence, violate NEC, the effective EoS parameter needs to be in the phantom area $w_{effective}<-1$ around the bouncing epoch. NEC, or null energy condition, is:
\begin{equation}
NEC \iff \rho_{Total} + P_{Total} \ge 0
\label{NEC}
\end{equation}
Therefore, from Eqs.\ref{total density}, \ref{2nd-Fried}, \ref{Hbounce} and \ref{NEC}, we can get the expression of NEC and plotted it Fig. \ref{Fig3_NEC}(Middle Panel);
\begin{equation}
\begin{array}{c}
NEC=\rho_{m_0} \left(A+B t^2\right)^{-3 \nu }-\frac{4 B \nu  \left(A+B t^2 (-1+3 \nu )\right)}{\left(A+B t^2\right)^2}\\+\frac{3 \left(\tau
_0+\left(C_1 \left(A+B t^2\right)^{\nu }-t \left(1+\frac{B t^2}{A}\right)^{\nu } {}_2F_1\left[\frac{1}{2},\nu ,\frac{3}{2},-\frac{B
t^2}{A}\right]\right) \tau _1+\left(C_1 \left(A+B t^2\right)^{\nu }-t \left(1+\frac{B t^2}{A}\right)^{\nu } {}_2F_1\left[\frac{1}{2},\nu
,\frac{3}{2},-\frac{B t^2}{A}\right]\right)^2 \tau _2\right){}^2}{\left(C_1 \left(A+B t^2\right)^{\nu }-t \left(1+\frac{B t^2}{A}\right)^{\nu
} {}_2F_1\left[\frac{1}{2},\nu ,\frac{3}{2},-\frac{B t^2}{A}\right]\right)^2}.
\end{array}
\label{NEC}
\end{equation}
We have now examined the stability of our model to see if it adequately represents the expanding universe with the background fluid NOGHDE. In this case, we calculated the square speed of sound \cite{squarespeed1,squarespeed2} in order to establish the stability. In order to ensure the model's stability, $V_s^2=\frac{dP_{effective}}{{d\rho_{effective}}} \ge 0$ is necessary. If the value is less than zero, the model is unstable. In Fig. \ref{Fig4_Vsquare}(right panel), we have illustrated the stability of our viscous non-interacting NOGHDE dark fluid model in bouncing condition through the square speed of sound against modest perturbations.
Here, we presented the total energy density ($\rho_{Total}$) in Fig. \ref{Fig2_rho} for the non-interacting bouncing case. In Fig. \ref{Fig2_rho}, the universe's necessary growth throughout time is obvious. In Fig. \ref{Fig2_rho}, the effective NOGHDE dark fluid increases in the post-bounce scenario due to expansion, decreases in the pre-bounce scenario due to contraction, and achieves its minimum value at $t = 0$ during the bounce. This figure illustrates a sharp rise in density caused by bounce inflation, which is consistent with the inflationary hypothesis of the universe. This is followed by a stable density for the course of the universe's subsequent periods. The symmetrical shape of the diagram encircling the bouncing point is explained by the presence of $t^2$ in the corresponding  Eq.(\ref{total density}).  The energy density is evolving steadily and continuously in the the background. The bounce is confirmed at $t = 0$. For a range of values of $\tau_0,\tau_1,\tau_2$, the NOGHDE parameters, we have displayed NEC and ${V_s}^2$ with respect to cosmic time $t$ in Figs. \ref{Fig3_NEC} (middile panel) and Fig. \ref{Fig4_Vsquare} (right panel). In order to create the diagrams for this section, we utilized the following parameters for the range of even values $\nu$ corresponding to the Red, Orange, Blue, and Green lines, respectively: $C_1 = 0.00006$, $\rho_{m_0} = 0.32$, $a_0 = 0.01$, $A = 1$, $B = 0.666, 1.5, 2.36, 4.5$, $c=0.58$. Each possibility is within suitable limits. In Fig. \ref{Fig3_NEC}(middile panel), it is shown that the reconstructed viscous single non-interacting NOGHDE supports bouncing behavior in the Einstein gravity framework during the cosmic evolution. There is an obvious bounce at $t=0$ where NEC is violated. The graph in Fig. \ref{Fig4_Vsquare} illustrates that the model is stable for all values of $\tau_0,\tau_1,\tau_2$ in early and later universe although near the bouncing point it is unstable.\\ 
Thus, we may obtain the formulation of the EoS parameter in a viscous, non-interacting, bouncing scenario from $\dot{\rho}_{DE}+3 H \rho_{DE}(1+w)=0$. 
Therefore, 
\begin{equation}
\begin{array}{c}
w=\frac{\eta_{11}}{\eta_{22}}
\end{array}
\label{w}
\end{equation}
\begin{equation*}
\begin{array}{c}
\eta_{11}=(\left(A+B t \left(t+C_1 \left(A+B t^2\right)^{\nu } \nu \right)-B t^2 \left(1+\frac{B t^2}{A}\right)^{\nu } \nu  {}_2F_1\left[\frac{1}{2},\nu
,\frac{3}{2},-\frac{B t^2}{A}\right]\right) \tau _0+\\
\left(C_1 \left(A+B t^2\right)^{\nu }-t \left(1+\frac{B t^2}{A}\right)^{\nu } {}_2F_1\left[\frac{1}{2},\nu
,\frac{3}{2},-\frac{B t^2}{A}\right]\right)^2 \\\left(3 B t \nu  \tau _1-\left(A+B t \left(t-5 C_1 \left(A+B t^2\right)^{\nu } \nu \right)+5
B t^2 \left(1+\frac{B t^2}{A}\right)^{\nu } \nu  {}_2F_1\left[\frac{1}{2},\nu ,\frac{3}{2},-\frac{B t^2}{A}\right]\right) \tau _2\right))
\end{array}
\end{equation*}
\begin{equation*}
\begin{array}{c}
\eta_{22}=(3 B t \nu  \left(-C_1 \left(A+B t^2\right)^{\nu }+t \left(1+\frac{B t^2}{A}\right)^{\nu } {}_2F_1\left[\frac{1}{2},\nu
,\frac{3}{2},-\frac{B t^2}{A}\right]\right)\\
(\tau _0+\left(C_1 \left(A+B t^2\right)^{\nu }-t \left(1+\frac{B t^2}{A}\right)^{\nu } {}_2F_1\left[\frac{1}{2},\nu
,\frac{3}{2},-\frac{B t^2}{A}\right]\right) (\tau _1+(C_1 \left(A+B t^2\right)^{\nu }\\-t \left(1+\frac{B t^2}{A}\right)^{\nu } {}_2F_1\left[\frac{1}{2},\nu
,\frac{3}{2},-\frac{B t^2}{A}\right]) \tau _2))
\end{array}
\end{equation*}
And, from Eq.(\ref{conservation_effective}), we can get the effective EoS parameter and plotted it in Fig \ref{Fig6_Weff}(right panel).
In Figs. \ref{Fig5_W} and \ref{Fig6_Weff}, both the EoS parameter and effective EoS parameter behaves like phantom scenario in pre- and post-bounce cases. Consequently, the evolution of the reconstructed EoS parameter (left panel) and effective EoS parameter (right panel) diverges at the bouncing point $t = 0$ as the denominator is zero. The Big Rip singularity is not prevented in pre- and post-bounce scenario in both the cases.

\section{Nojiri-Odintsov Generalized Holographic Dark Fluid-Driven Bounce Inflation}
In this section, we intend to carry out a detailed study on the bounce and inflationary cosmology, considering the driving fluid as the holographic fluid with NO cutoff. It may be noted that this is the most generalized version of holographic fluid. So far, holographic fluid has been considered mostly for investigating the late-time accelerated expansion of the universe. However, in the present case, we consider the early inflationary explanation and bounce scenario with NO holographic viscous fluid. Furthermore, the possibility of holographic realization followed by a bouncing scenario is already investigated. In view of the above, we will discuss the Hubble slow-roll approximation and correspondence with scalar field models in the subsequent subsections. We would like to mention that, bulk viscosity has been considered while dealing with the NO holographic fluid for inflation and bounce. The details are demonstrated in the subsequent sections.
\subsection{Hubble Slow-roll Approximation}
We will address a scenario where there is a transition to accelerated expansion at the end of a non-singular bounce, inspired by the studies \cite{bounceinflation9,bounceinflation10,bounceinflation11,bounceinflation12}. The original Friedmann equation for a generic inflationary scenario is $H^2=\frac{\rho_{inf}}{3}$, where $\rho_{inf}$ denotes the energy density of the (effective) fluid driving inflation, which can originate from a scalar field. Note that, as usual, we could not include the contributions from the radiation components. This work investigates the notion that the Nojiri-Odintsov generalized holographic dark energy and non-interacting pressure-less dark matter drive inflation. Since the growth of $H(t)$ is known in bouncing viscous NOGHDE scenario, the Hubble slow-roll parameters $\epsilon_n$ (where $n$ is a positive integer) can therefore be calculated \cite{slowroll1a,slowroll2a,slowroll3a,slowroll4a}. Consequently,
\begin{equation}
\epsilon_{n+1}=\frac{d~ln|\epsilon_n|}{dN}, 
\end{equation}
where, $N=ln(\frac{a}{a_{initial}})$ is the definition of the e-folding number $N$, with $\epsilon_0=\frac{H_{initial}}{H}$. Although we did not use the e-folding time scale in this present study, we have used the original cosmic time frame. In this case, $a_{initial}$ represents the scale factor at the start of inflation, corresponding to $H_{initial}$, the Hubble parameter. At this point, we may calculate the values of the inflationary observables, as stated in \cite{slowroll3a}: the tensor spectral index $n_T$, the tensor-to-scalar ratio $r$, and the scalar spectral index of the curvature perturbations $n_s$. The first three Hubble slow-roll parameters are-
\begin{equation}
\epsilon_1=\frac{-\dot{H}}{H^2},
\label{epsilon1}
\end{equation}
\begin{equation}
\epsilon_2=\frac{\ddot{H}}{H\dot{H}}-\frac{2\dot{H}}{H^2},
\label{epsilon2}
\end{equation}
and,
\begin{equation}
\epsilon_3=(\ddot{H}H-2\dot{H}^2)^{-1}
[\frac{H\dot{H}\dddot{H}-\ddot{H}(\dot{H}^2+H\ddot{H}}{H\dot{H}}-\frac{2\dot{H}}{H^2}(H\ddot{H}-2\dot{H}^2)].
\label{epsilon3}
\end{equation}
Again,
\begin{equation}
r\equiv16\epsilon_1,~~\text{and}~~ n_s\equiv 1-2\epsilon_1-2\epsilon_2~.
\label{r-ns}
\end{equation}
Then,
\begin{equation}
\alpha_s\equiv -2\epsilon_1\epsilon_2-\epsilon_2\epsilon_3
\label{alphas}
\end{equation}
\begin{equation}
n_T\equiv -2\epsilon_1
\label{nT}
\end{equation}
To check inflationary expansion of the background viscous non-interacting NOGHDE fluid following the non-singular bounce, we have put the expression of total energy density in Eq.(\ref{epsilon1}) by 1st Friedmann equation. Therefore,
\begin{equation}
\begin{array}{c}
\epsilon_1=-\frac{6 B \left(A-B t^2\right) \nu }{\left(A+B t^2\right)^2 (\rho_{m_0} \left(A+B t^2\right)^{-3 \nu }+\frac{M_{11}}{M_{22}}})
\end{array}
\label{epsilon1-NO}
\end{equation}
where,
\begin{equation*}
\begin{array}{c}
M_{11}=3
(\tau _0+\left(C_1 \left(A+B t^2\right)^{\nu }-t \left(1+\frac{B t^2}{A}\right)^{\nu } {}_2F_1\left[\frac{1}{2},\nu ,\frac{3}{2},-\frac{B
t^2}{A}\right]\right)\\ \left(\tau _1+\left(C_1 \left(A+B t^2\right)^{\nu }-t \left(1+\frac{B t^2}{A}\right)^{\nu } {}_2F_1\left[\frac{1}{2},\nu
,\frac{3}{2},-\frac{B t^2}{A}\right]\right) \tau _2\right)){}^2 
\end{array}
\end{equation*}
and,
\begin{equation*}
M_{22}=\left(C_1 \left(A+B t^2\right)^{\nu }-t \left(1+\frac{B t^2}{A}\right)^{\nu
} {}_2F_1\left[\frac{1}{2},\nu ,\frac{3}{2},-\frac{B t^2}{A}\right]\right)^2
\end{equation*}
We can now evaluate $\epsilon_2$ from Eqs.(\ref{epsilon1}), (\ref{epsilon1-NO}), (\ref{epsilon2}) and (\ref{1st-Fried}) and from these $\epsilon_1~\text{and}~ \epsilon_2$ induced by viscous non-interacting NOGHDE fluid, we can easily calculate the tensor-to-scalar ratio $r$, and the scalar spectral index of the curvature perturbations $n_s$ by Eq.(\ref{r-ns}).
\begin{figure}
\centering
\begin{minipage}{.5\textwidth}
  \centering
\includegraphics[width=.9\linewidth]{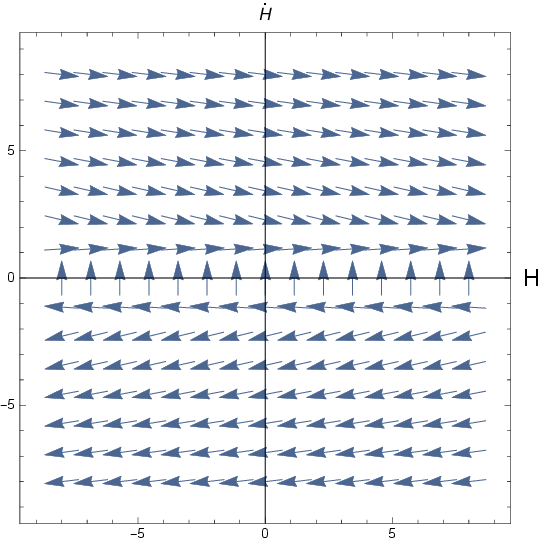}
\caption{Plotting of phase-space diagram in viscous non-interacting bouncing NOGHDE scenario.}
\label{Fig7_phase-space}
\end{minipage}%
\centering
\begin{minipage}{.5\textwidth}
  \centering
\includegraphics[width=.9\linewidth]{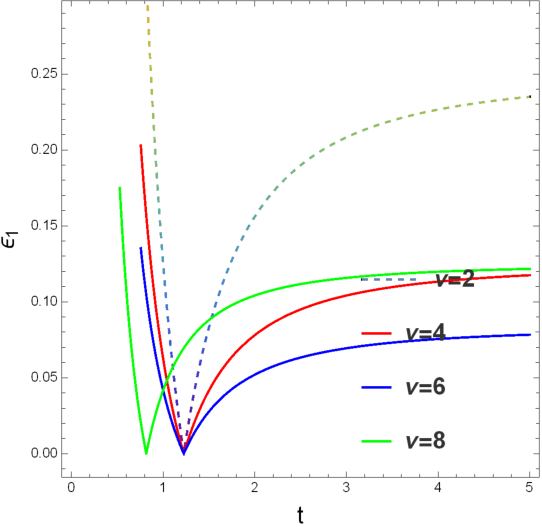}
\caption{Evolution of reconstructed first Hubble flow parameter $\epsilon_1$ with respect to cosmic time $t$ in viscous non-interacting bouncing NOGHDE scenario.}
\label{Fig8_epsilon1}
\end{minipage}%
\end{figure}
In this subsection, we first plotted a phase-space diagram in Fig. \ref{Fig7_phase-space}. This diagram shows the four quadrants. Observing the minutes of the diagram we can say that before as well as after the bounce $\dot{H}$ is at positive level. This indicates that in a combination of negative $H$ and positive $\dot{H}$ is apparent in the pre-bounce scenario. This is interpreted to lead to pre-bounce contraction in the cosmological setting of NO holographic viscous dark fluid. On the other hand, as we observed both $H~\text{and}~\dot{H}$ to be positive in the post-bounce scenario, the post-bounce expansion is attainable under the said cosmological framework. At the turn-around point, we have $H=0$ without any discontinuity in the evolutionary pattern of the Hubble parameter. This indicates a non-singular transition from pre- to post-bounce universe. In Fig. \ref{Fig8_epsilon1} the evolution of the first Hubble flow parameter $\epsilon_1$ is plotted with respect to cosmic time for the scenario elaborated above. It is observed that, in the post-bounce scenario, $\epsilon<<1$. Therefore, we can see, the realisation of inflation by the viscous non-interacting NOGHDE fluid following the non-singular bounce.

\subsection{Scalar Field Correspondence}
In this subsection, we discussed the scalar field correspondence of the generalized holographic dark fluid with Nojiri-Odintsov cutoff with three scalar field models: canonical, tachyon, and k-essence. Details of the cosmological consequences are appended in the following subsections. These subsections aim to check the realization of inflation in inflationary expansion following a bounce scenario under the cosmological settings of NO generalized holographic fluid.
\subsubsection{Canonical Scalar Field}
\begin{figure}
\centering
\begin{minipage}{.5\textwidth}
  \centering
\includegraphics[width=.9\linewidth]{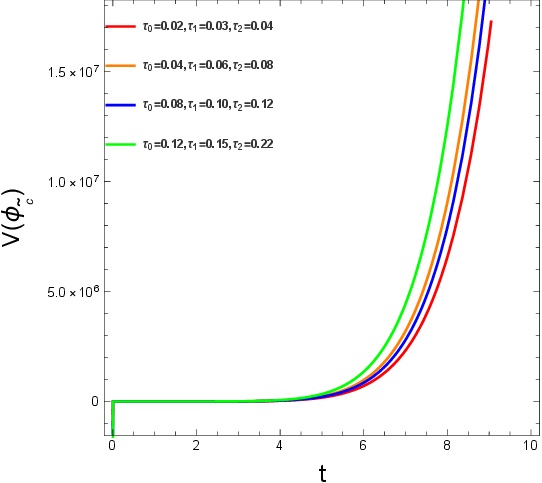}
\caption{Evolution of reconstructed canonical potential $V\left(\phi _{\underset{c}{\sim }}\right)$ with respect to cosmic time $t$ in viscous non-interacting bouncing NOGHDE scenario.}
\label{Fig9_Vphi-Canonical}
\end{minipage}%
\begin{minipage}{.5\textwidth}
  \centering
\includegraphics[width=.9\linewidth]{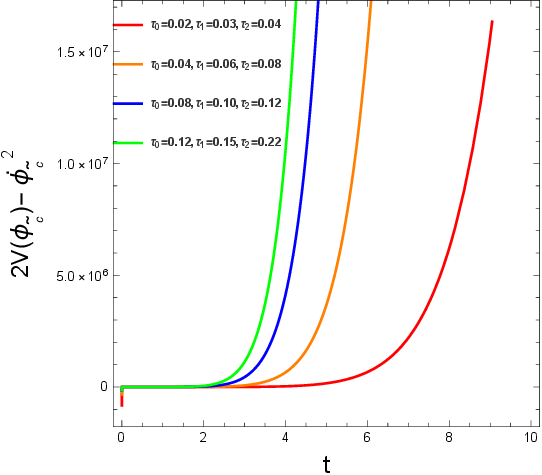}
\caption{Evolution of reconstructed $2V\left(\phi _{\underset{c}{\sim }}\right)-\dot{\phi}_{\tilde{c}}^2$ with respect to cosmic time $t$ in viscous non-interacting bouncing NOGHDE scenario.}
\label{Fig10_2V-Canonical}
\end{minipage}%
\end{figure}

In this subsection, we show that inflationary behavior can be described regarding scalar field dynamics using the NOGHDE approach. \cite{DE8} presents a work that shows how a scalar field and a DE correspond. Scalar field dynamics has been fundamental to current cosmology since inflation, leading to a paradigm shift combining high-energy physics and cosmology. We first look at the connection between NOGHDE and the canonical scalar field. The pressure and energy density of the canonical scalar field are as follows, according to \cite{Canonical}:
\begin{equation}
\rho_{\phi_{\tilde{c}}}=\frac{1}{2}\dot{\phi}_{\tilde{c}}^2+V(\phi_{\tilde{c}})
\label{rhophi-canonical}
\end{equation}
\begin{equation}
P_{\phi_{\tilde{c}}}=\frac{1}{2}\dot{\phi}_{\tilde{c}}^2-V(\phi_{\tilde{c}})
\label{Pphi-canonical}
\end{equation}
where, $\phi_{\tilde{c}}$ be the canonical scalar field. Therefore, potential and scalar field can be written as:
\begin{equation}
V(\phi_{\tilde{c}})=\frac{1}{2}\rho_{effective}(1-w_{effectice})
\label{Vphi-canonical}
\end{equation}
and, \begin{equation}
\dot{\phi}_{\tilde{c}}^2=\rho_{effective}(1+w_{effective})
\label{phidotsquarecanonical}
\end{equation}
We obtain the potential $V(\phi_{\tilde{c}})$ by substituting $\rho_{NOGHDE}$ and $w_{NOGHDE}$ in Eq.(\ref{Vphi-canonical}) by total fluid density of NOGHDE $\rho_{Total}$ and effective EoS parameter $w_{eff}$ of bouncing viscous non-interacting scenario from Eq.(\ref{total density}) and effective EoS parameter respectively. Therefore,
\begin{equation}
\begin{array}{c}
V\left(\phi _{\underset{c}{\sim }}\right)=\frac{1}{2} \left(\rho_{m_0} \left(A+B t^2\right)^{-3 \nu }+\frac{\eta_{44}}{\eta_{55}}\right)
\left(1+\frac{12 B^2 t^2 \nu ^2 \xi }{\left(A+B t^2\right)^2}-\frac{M_{33}}{M_{44}M_{55}}\right)
\end{array}
\label{Vphi-canonical}
\end{equation}
where, 
\begin{equation}
\begin{array}{c}
\eta_{44}=3 (\tau _0+\left(C_1 \left(A+B t^2\right)^{\nu }-t \left(1+\frac{B t^2}{A}\right)^{\nu } {}_2F_1\left[\frac{1}{2},\nu
,\frac{3}{2},-\frac{B t^2}{A}\right]\right)\\ \left(\tau _1+\left(C_1 \left(A+B t^2\right)^{\nu }-t \left(1+\frac{B t^2}{A}\right)^{\nu } {}_2F_1\left[\frac{1}{2},\nu
,\frac{3}{2},-\frac{B t^2}{A}\right]\right) \tau _2\right))^2, 
\end{array}
\end{equation}
\begin{equation}
\begin{array}{c}
\eta_{55}=\left(C_1 \left(A+B t^2\right)^{\nu }-t \left(1+\frac{B t^2}{A}\right)^{\nu } {}_2F_1\left[\frac{1}{2},\nu
,\frac{3}{2},-\frac{B t^2}{A}\right]\right)^2,
\end{array}
\end{equation}
\begin{equation}
\begin{array}{c}
M_{33}=((\tau _0+\left(C_1 \left(A+B t^2\right)^{\nu }-t \left(1+\frac{B t^2}{A}\right)^{\nu } {}_2F_1\left[\frac{1}{2},\nu
,\frac{3}{2},-\frac{B t^2}{A}\right]\right)\\ (\tau _1+(C_1 \left(A+B t^2\right)^{\nu }-t \left(1+\frac{B t^2}{A}\right)^{\nu } {}_2F_1\left[\frac{1}{2},\nu
,\frac{3}{2},-\frac{B t^2}{A}\right]) \tau _2))\\ 
((A+B t \left(t+C_1 \left(A+B t^2\right)^{\nu } \nu \right)-B t^2 \left(1+\frac{B t^2}{A}\right)^{\nu } \nu  {}_2F_1\left[\frac{1}{2},\nu
,\frac{3}{2},-\frac{B t^2}{A}\right]) \tau _0+
(C_1 (A+B t^2)^{\nu }\\-t (1+\frac{B t^2}{A})^{\nu } {}_2F_1\left[\frac{1}{2},\nu
,\frac{3}{2},-\frac{B t^2}{A}\right])^2 (3 B t \nu  \tau _1-(A+B t \left(t-5 C_1 \left(A+B t^2\right)^{\nu } \nu \right)\\+5
B t^2 \left(1+\frac{B t^2}{A}\right)^{\nu } \nu  {}_2F_1\left[\frac{1}{2},\nu ,\frac{3}{2},-\frac{B t^2}{A}\right]) \tau _2))),
\end{array}
\end{equation}
\begin{equation}
\begin{array}{c}
M_{44}=(B t \nu  \left(-C_1 \left(A+B t^2\right)^{\nu }+t \left(1+\frac{B t^2}{A}\right)^{\nu } {}_2F_1\left[\frac{1}{2},\nu
,\frac{3}{2},-\frac{B t^2}{A}\right]\right)^3),
\end{array}
\end{equation}
\begin{equation}
\begin{array}{c}
M_{55}=(\rho_{m_0} \left(A+B t^2\right)^{-3 \nu }+\\\frac{3 \left(\tau _0+\left(C_1 \left(A+B t^2\right)^{\nu }-t \left(1+\frac{B
t^2}{A}\right)^{\nu } {}_2F_1\left[\frac{1}{2},\nu ,\frac{3}{2},-\frac{B t^2}{A}\right]\right) \left(\tau _1+\left(C_1 \left(A+B
t^2\right)^{\nu }-t \left(1+\frac{B t^2}{A}\right)^{\nu } {}_2F_1\left[\frac{1}{2},\nu ,\frac{3}{2},-\frac{B t^2}{A}\right]\right)
\tau _2\right)\right){}^2}{\left(C_1 \left(A+B t^2\right)^{\nu }-t \left(1+\frac{B t^2}{A}\right)^{\nu } {}_2F_1\left[\frac{1}{2},\nu
,\frac{3}{2},-\frac{B t^2}{A}\right]\right)^2}).
\end{array}
\end{equation}
The potential is reconstructed in Eq.(\ref{Vphi-canonical}) for the canonical scalar field model. The reconstructed potential as obtained in Eq.(\ref{Vphi-canonical}) is depicted diagrammatically in Fig.\ref{Fig9_Vphi-Canonical} where the potential is plotted as a function of the canonical scalar field against cosmic time $t$. The reconstructed potential is observed to be positive level with an increasing pattern. Furthermore, in Fig.\ref{Fig10_2V-Canonical} we have plotted $2V(\phi_{\tilde{c}})-\dot{\phi}_{\tilde{c}}^2$ against the cosmic time $t$. It is clear that $2V(\phi_{\tilde{c}})-\dot{\phi}_{\tilde{c}}^2$ is greater than zero which indicates the realization of inflationary expansion. Thus it can be interpreted that a bounce-inflation is attainable with viscous NO holographic fluid as a correspondence with canonical scalar field is considered.
\subsubsection{Tachyon Scalar Field}
Moreover, tachyon fields have been postulated as possible dark energy candidates \cite{Tachyon}.
Since tachyons are hypothetical mass scalar fields, they may always move faster than the speed of light. Despite this apparently paradoxical aspect, models of tachyonic dark energy have been developed where the tachyon field causes faster expansion. Since these models involve complicated dynamics, a careful analysis is required to ensure they make coherence in light of theoretical constraints and observable data. This section aims to investigate the circumstances in which the related NOGHDE acts as a tachyonic field. Our work employs an integrated approach to look for similarities, underlying concepts, and potential connections between canonical and tachyonic scalar field models. With the non-singular Big Bounce  acting as the universe's origin in this case, establishing such a framework may give a more comprehensive understanding of the related NOGHDE's role in the universe's accelerated expansion. The tachyon field's pressure and energy density are determined by:
\begin{equation}
\rho_{\phi_{\tilde{T}}}=\frac{V(\phi_{\tilde{T}})}{\sqrt{1-\dot{\phi}_{\tilde{T}}^2}}
\label{rhophi-tachyon}
\end{equation}
\begin{equation}
P_{\phi_{\tilde{T}}}=-V(\phi_{\tilde{T}}){\sqrt{1-\dot{\phi}_{\tilde{T}}^2}}
\label{Pphi-tachyon}
\end{equation}
where, tachyonic scalar field is denoted by $\phi_{\tilde{T}}$. We induce the effective pressure and effective energy densities of this viscous non-interacting NOGHDE model in order to determine an appropriate potential for the tachyonic field that illustrates bounce inflationary behavior. The result is:
\begin{equation}
V(\phi_{\tilde{T}})=\rho_{Total}\sqrt{1-\dot{\phi}_{\tilde{T}}^2}
\label{Vphi-tachyon}
\end{equation}
where, \begin{equation}
\dot{\phi}_{\tilde{T}}^2=1+w_{eff}
\label{phidot-tachyon}
\end{equation}
\begin{figure}
\centering
\begin{minipage}{.5\textwidth}
  \centering
\includegraphics[width=.9\linewidth]{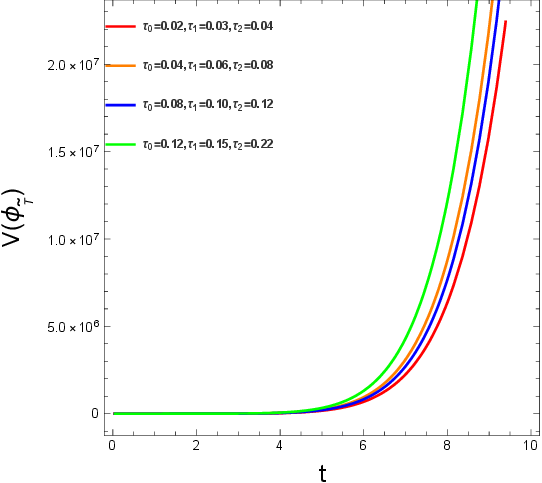}
\caption{Evolution of reconstructed tachyonic potential $V\left(\phi _{\underset{T}{\sim }}\right)$ with respect to cosmic time $t$ in viscous non-interacting bouncing NOGHDE scenario.}
\label{Fig11_Vphi-Tachyon}
\end{minipage}%
\begin{minipage}{.5\textwidth}
  \centering
\includegraphics[width=.9\linewidth]{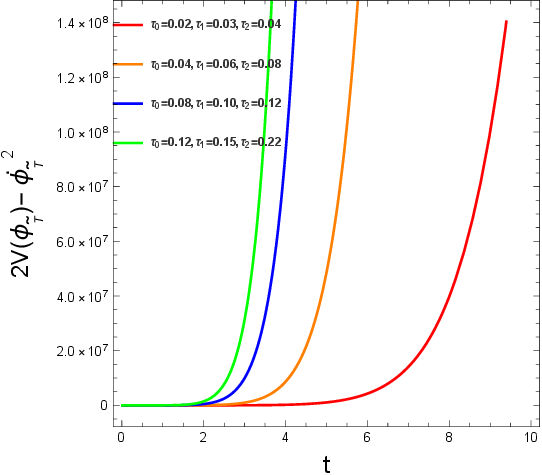}
\caption{Evolution of reconstructed  $2V\left(\phi _{\underset{T}{\sim }}\right)-\dot{\phi}_{\tilde{T}}^2$ with respect to cosmic time $t$ in viscous non-interacting bouncing NOGHDE scenario.}
\label{Fig12_2V-Tachyon}
\end{minipage}%
\end{figure}
Like the canonical scalar field the possibility of bounce inflation is noticeable in the case of NOGHDE. Please see Fig.\ref{Fig11_Vphi-Tachyon} and Fig. \ref{Fig12_2V-Tachyon} for the reconstructed $V(\phi_{\tilde{T}})$ and $2 V(\phi_{\tilde{T}})- \dot{\phi}_{\tilde{T}}^2$. These graphs illustrate how potential rises as time values do. The observations align with this conclusion that the realization of inflationary expansion is established in post-bounce scenario. Fig. \ref{Fig12_2V-Tachyon} plots $2V-\dot{\phi}_{T}^2$ vs time; this is consistent with the observational data, which indicates that $2V>>\dot{\phi}_{T}^2$ i.e inflationary expansion.

\subsubsection{K-essence Scalar Field}
\begin{figure}
\centering
\begin{minipage}{.5\textwidth}
  \centering
\includegraphics[width=.9\linewidth]{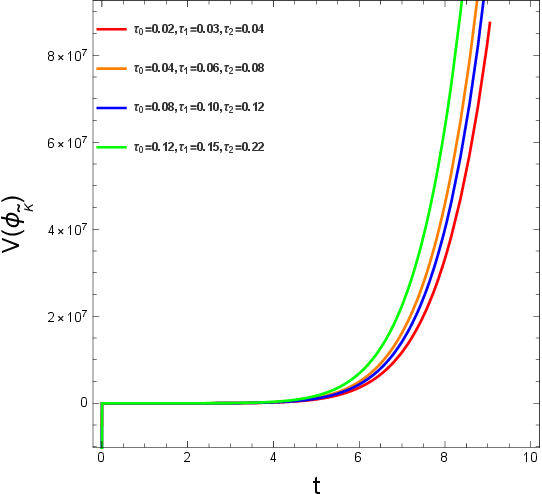}
\caption{Evolution of reconstructed K-essence potential $V(\phi_{\tilde{K}})$ with respect to cosmic time $t$ in viscous non-interacting bouncing NOGHDE scenario.}
\label{Fig13_Vphi-K-essence}
\end{minipage}%
\begin{minipage}{.5\textwidth}
  \centering
\includegraphics[width=.9\linewidth]{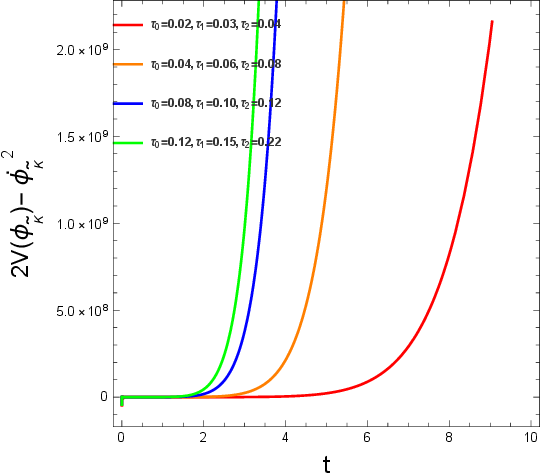}
\caption{Evolution of reconstructed  $2V(\phi_{\tilde{K}})-\dot{\phi}_{\tilde{K}}^2$ with respect to cosmic time $t$ in viscous non-interacting bouncing NOGHDE scenario.}
\label{Fig14_2V-K-essence}
\end{minipage}%
\end{figure}

\begin{figure}
\centering
\begin{minipage}{.5\textwidth}
  \centering
\includegraphics[width=.9\linewidth]{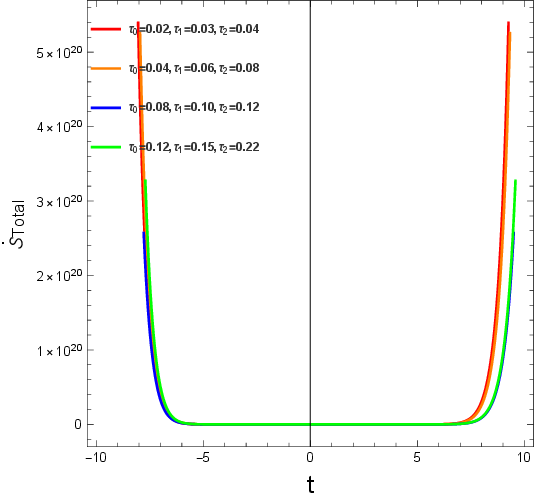}
\caption{Evolution of time derivative of total entropy $(\dot{S})$ with time $(t)$, using Bekenstein entropy as entropy at the future event horizon  with the background fluid as reconstructed generalized holographic dark fluid with Nojiri-Odintsov cut-off in viscous non-interacting bouncing scenario.}
\label{Fig15_Sdot}
\end{minipage}%
\centering
\begin{minipage}{.5\textwidth}
  \centering
\includegraphics[width=.9\linewidth]{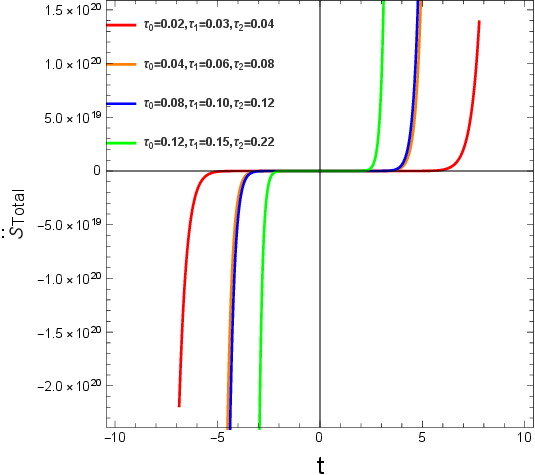}
\caption{Plot of thermal equilibrium $\ddot{\mathcal{S}}$ with time $(t)$, using Bekenstein entropy as entropy at the future event horizon  with the background fluid as reconstructed generalized holographic dark fluid with Nojiri-Odintsov cut-off in viscous non-interacting bouncing scenario.}
\label{Fig16_Sddot}
\end{minipage}%
\end{figure}

According to the k-essence model\cite{k-essence}, the energy density and pressure are as follows:
\begin{equation}
\rho_{\tilde{K}}=V(\phi_{\tilde{K}})(-\mathcal{M}+3 \mathcal{M}^2)
\label{rhophi-K-essence}
\end{equation}
and, \begin{equation}
P_{\tilde{K}}=V(\phi_{\tilde{K}})(-\mathcal{M}+ \mathcal{M}^2);
\label{P-phi-K-essence}
\end{equation}
where, $\mathcal{M}=\frac{1}{2}\dot{\phi}_{\tilde{K}}^2.$
The EoS parameter can be written as, 
\begin{equation}
w_{\tilde{K}}=\frac{\mathcal{M}-1}{3 \mathcal{M}-1};
\label{w-phi-K-essence}
\end{equation}
wherein $\mathcal{M}$ experienced the accelerated expansion of the universe during the span $(\frac{1}{3},\frac{2}{3})$. We may derive the scalar field $\dot{\phi}_{\tilde{K}}^2$ and the potential $V(\phi_{\tilde{K}})$ for this K-essence scalar field circumstance by taking $w_{\tilde{K}} = w_{eff}$ in this viscous non-interacting NOGHDE scenario. However, we have observed that the pattern of potential $V(\phi_{\tilde{K}})$ in Fig.\ref{Fig13_Vphi-K-essence} is monotonically increasing as previous cases and $2V>>\dot{\phi}_{K}^2$ is noticeable in Fig.\ref{Fig14_2V-K-essence} which supports inflationary expansion in post- bounce scenario. 
\textcolor{black}{\section{Validation of GSL in Viscous Bouncing NOGHDE Universe}}
We next investigate the validity of the generalized second law of thermodynamics after determining the total entropy change rate in this section. The gravitational theory of thermodynamic analysis is a well-known field of fascinating cosmology research, and cosmic horizons share many of the same thermodynamic characteristics as black holes\cite{Odintsov5,holographic principle3,thermodynamic1,thermodynamic3,thermodynamic4,thermodynamic5,thermodynamic6,thermodynamic7,thermodynamic8,thermodynamic9,thermodynamic10,thermodynamic11,thermodynamic12}. Furthermore, the first rule of thermodynamics, which holds at a black hole horizon, may also be inferred from the first Friedmann equation in the FRW universe when the universe is constrained by an future event horizon. This provides a strong argument for using the future event horizon as the cosmic horizon when examining the thermodynamic properties of any cosmological model. Here, we have chosen viscous bouncing non-interacting generalized holographic dark fluid with Nojiri-Odintsov cut-off for this thermodynamic calculation. In light of the aforementioned justifications, we have regarded the universe as a thermodynamic system here, enclosed by the cosmic future event horizon of radius. After the thermodynamic properties of the black hole were clarified \cite{Odintsov5,holographic principle2,thermodynamic1}, it has been claimed that the entropy of the black hole is proportional to the area $A$ of the horizon, which is
\begin{equation}
\mathcal{S}=\frac{A}{4G},~\text{where},~A=4\pi \mathcal{R}_h^2 .
\end{equation}
The relationship between thermodynamics and gravity has been explained in numerous recent studies. The horizon radius is denoted by $\mathcal{R}_h$, and we work in units where $h = k_B = c = 1$ \cite{thermodynamic3,thermodynamic4}. This is known as the Bekenstein-Hawking entropy. The first rule of thermodynamics, which holds in a black hole horizon, may also be found using the first Friedmann equation in the FRW universe when an apparent horizon constrains the universe. Bekenstein\cite{holographic principle2} proposed in 1973 that the thermodynamics and future event horizon of a black hole are related, i.e., the event horizon of a black hole is a measure of its entropy. We have extended this idea to cosmological model horizons, where each horizon equals an entropy. Thus, in its generalized form, the total of all horizon-related time derivatives of entropies plus the time derivative normal entropy must be positive, with respect to the second law of thermodynamics. This means that the time derivative of total of entropies must be positive. The distance that light travels from the present to infinity, or the future event horizon, is determined by Eq.(\ref{future-event})

where, $a(t)$ denotes the choice of scale factor. Let us consider the overall entropy $(S)$ will be $S=S_h+S_f$  where, $S_f$ and $S_h$ represent the entropies of the fluid and the horizon that contains it, respectively. The following relations, which are in accordance with the laws of thermodynamics, must be satisfied by $S$, like by any solitary macroscopic system. Therefore, by second law of thermodynamics(GSL) and thermodynamic equilibrium (TE),
\begin{equation}
\begin{array}{c}
\dot{S}=\frac{dS}{dt}\ge0~~and~~\ddot{S}=\frac{d^2S}{dt^2}<0.
\end{array}
\label{sdot}
\end{equation}
To verify the validity of GSL, we have considered both viscous NOGHDE and dark matter in a non-interacting scenario. We evaluate the viability of GSL using the assumption that the future event horizon and the dark sectors are in equilibrium, i.e., that the temperature of the horizon is the same for dark energy and dark matter. The event horizon's temperature is
\begin{equation}
\begin{array}{c}
T_E=\frac{1}{2 \pi \mathcal{R}_h}
\end{array}
\label{event-temperature}
\end{equation}
As previously mentioned, we considered the viscous NOGHDE and DM as the components of the background fluid, then we can write,
\begin{equation}
S_f=S_{NOGHDE}+S_{DM}
\label{entropy-fluid}
\end{equation}
where, $T$ is taken to be the temperature of this single fluid inside the horizon, and $S_{NOGHDE}$ and $S_{DM}$ denote the entropies of the non-interacting fluid of NOGHDE and DM, respectively. Consequently, the first law of thermodynamics $(TdS = dE + PdV)$ can be written as follows for the individual matter contents:
\begin{equation}
TdS_{NOGHDE} = dE_{NOGHDE} + P_{NOGHDE} dV,
\label{TdS_{NOGHDE}}
\end{equation}
\begin{equation}
TdS_{DM} = dE_{DM} + P_{DM}dV = dE_{DM},
\label{TdS_{DM}}
\end{equation}
where the horizon volume is expressed as $V = \frac{4}{3} \pi \mathcal{R}_h^3$. Again, $E_{NOGHDE}=\frac{4}{3} \pi \mathcal{R}_h^3 \rho_{NOGHDE}$ and $E_{DM}=\frac{4}{3} \pi \mathcal{R}_h^3\rho_{DM}$ represents the internal energies of NOGHDE and pressure-less DM.
Therefore, by differentiating Eq.(\ref{TdS_{NOGHDE}}) and Eq.(\ref{TdS_{DM}}), we can get $\dot{S}_{NOGHDE}$ and $\dot{S}_{DM}$ respectively. Furthermore, it is important to remember that in this specific situation, the geometry would deform due to the energy flow if the fluid temperture $T$ differs from the event horizon temperture $T_h$. Therefore, the entropy inside this non-interacting NOGHDE and DM will be:
\begin{equation}
\begin{array}{c}
\dot{S}_f=\dot{S}_{NOGHDE}+\dot{S}_{DM}
\end{array}
\label{entropy-fluid}
\end{equation}
Again, entropy of the horizon can be defined as:
\begin{equation}
\begin{array}{c}
S_h=\frac{A}{4G}, ~\text{where,}~A=4\pi \mathcal{R}_h^3
\end{array}
\label{entropy-horizon}
\end{equation}
Here, $A=4\pi \mathcal{R}_h^3$ be the surface area and $\mathcal{R}_h$ be the radius of the future event horizon which we got from Eq.(\ref{Rh}). Therefore, by differentiating Eq.(\ref{entropy-fluid}), we can arrive at the expression:
\begin{equation}
\begin{array}{c}
\dot{S}_f= \dot{S}_{NOGHDE}+\dot{S}_{DM}
\end{array}
\label{dot-entropy-fluid}
\end{equation}
Therefore,
\begin{equation}
\dot{S}=\dot{S}_h+\dot{S}_{NOGHDE}+\dot{S}_{DM}
\label{time derivative of total entropy}
\end{equation}
Hence, by replacing $\mathcal{R}_h$ from Eq.(\ref{Rh}), $\rho_{m}$ from Eq.(\ref{density darkmatter}), $\rho_{Total}$ and $P$ of viscous non-interacting bouncing NOGHDE in Eqs.\ref{entropy-fluid},\ref{entropy-horizon}, we can get the time derivative of total entropy in bouncing non-interacting scenario. It is plotted in Fig. \ref{Fig15_Sdot}. Fig. \ref{Fig15_Sdot} is consistent with the time derivative of total entropy as it satisfies the inequality $\dot{S}\ge 0$ in both pre- and post-bounce scenarios. Thermal equilibrium(TE) is also established in post-bounce scenario for $\ddot{S} \leq 0$.

\section{Conclusion}
In this study, we implemented bounce cosmology using the holographic principle with the specific cutoff introduced by Nojiri and Odintsov\cite{NO}.  With its intriguing phenomenology, the holographic bounce has garnered a lot of curiosity. We considered the infrared cutoff to be equal to the future-event horizon and calculated the analytic cutoff in order to study cosmological models with a bouncing scale factor. In this case, we computed the energy conservation equation in holographic language, assuming a bulk viscosity for the cosmic fluid. 
As a result, we looked into holographic bounce cosmology in viscosity and demonstrated how it showed equivalence with the specific cutoff introduced by Nojiri and Odintsov. Though inflation theories have made great progress, the singularity problem has plagued them. Phenomenologically, the singularity can be avoided by considering the fact that a non-singular bounce precedes inflation. The bounce raises the possibility that it was contracting from a big volume before the Universe started to expand. Apart from providing a rationale for the enigmatic subject of what our Universe looked like before inflation, the bounce scenario might potentially incorporate some new elements into the early stages of our Universe and make predictions that could be discovered in subsequent observations. 

In the work of \cite{holographic bounce}, the authors examined the bounce realization while also employing the holographic concept at an early stage. Because they provide a potential solution to the cosmic singularity problem and an alternative to inflation, nonsingular bouncing cosmologies are of great interest. Our goal for this work was to establish a link between non-singular bounce inflation and NOGHDE.
The cosmological inflationary dynamics of scalar fields are also examined in this paper. Nojiri-Odintsov generalized holographic dark energy (NOGHDE) is chosen in the viscous non-interacting scenario for this case. In this study, we have demonstrated, the possibility of unifying the early-time and late-time universe based on bounce-inflationary cosmology when the generalized holographic dark energy with a Nojiri-Odintsov cut-off (which we have only focused on throughout this paper) was considered. Dissipative processes including bulk viscosity, shear viscosity, and heat transport majorly impact cosmic expansion.  A cosmic fluid may lose its local thermodynamic equilibrium and acquire bulk viscosity if it expands (or contracts) too quickly. We have therefore investigated bouncing cosmology in the context of viscous NOGHE in a non-interacting setting. By adopting the Big-bounce cosmology, a single NOGHDE plays a crucial role in realizing inflation in this viscous world. The expressions for the Hubble parameter $H$, evolution of reconstructed Nojiri-Odintsov generalized dark fluid density $(\rho_{Total})$, and violation of null-energy condition are plotted in Figs. \ref{Fig1_H}, \ref{Fig2_rho} and \ref{Fig3_NEC} respectively. In  Fig.(\ref{Fig3_NEC}), this generalized version of the holographic dark fluid model has been realized big-bounce in the framework of Einstein gravity, which violates the null energy condition (NEC) $P_{effective} + \rho_{effective} \geq 0$ for all even values of $\nu$. In  Fig.(\ref{Fig2_rho}), we can see, the bouncing point appeared at $t = 0$, with $H>0~\text{and} \dot{H}>0$ in the post-bounce scenario, although $H<0~\text{with}~\dot{H}>0$ in pre-bounce scenario. Fig. \ref{Fig4_Vsquare} shows that, for all positive values of $\tau_0, \tau_1, \tau_2$ in both early and late universe, our model is stable against small perturbations although it is showing instability during bouncing era of the universe. As illustrated in Figs. \ref{Fig5_W} and \ref{Fig6_Weff}, the growth of the reconstructed EoS parameter (left panel) and effective EoS parameter(right panel) diverges due to viscous NOGHDE in non-interacting scenario. The EoS parameters lie in phantom boundary in pre- and post-bounce instances. 

The holographic principle is widely used to explain late time acceleration in literatures. Inspired by this, we have used the theory in the early time inflationary scenario here. The concept states that the energy density is proportional to the inverse of length squared. Since the length scale is expected to be very tiny during the inflation, the energy density created should be adequate to drive the inflation. Since holographic dark energy is known to originate from entropy, altering the entropy law will alter the characteristics of the DE. We have studied an inflationary scenario characterized by a universe filled of visocus non-interacting NOGHDE, with $\nu$ being positive even integers. Several analytical solutions were found for the model in the non-singular bouncing scenario: the tensor-to-scalar ratio, scalar spectral index, and slow-roll parameters. Finally, a potential correspondence between the NOGHDE and scalar fields is explored. The canonical scalar field, the Tachyonic field, and the K-essence are used for this. The three fields' potential can be plotted to examine its evolution. It is evident that the trend is consistent with the observational data. In a flat Friedmann universe, we have constructed a model that can stably realize a transition from contraction to expansion by Fig. \ref{Fig7_phase-space} without deviating from the weak-gravity regime. For the bounce to be considered sound, the evolution must be devoid of gradient instabilities and the perturbations must remain ghost-free. More accurately, we have derived a series of necessary criteria on the form of bounce inflation theory which guarantee the stability of the evolution at the turnaround point. In Fig. \ref{Fig8_epsilon1}, we have shown that rapid inflation can be easily realized in a post-bounce scenario. But we have not yet succeeded in developing a model that explains how expansion has occurred throughout history, both in the far past and far future. Like most non-singular bouncing cosmologies, the model here avoids the cosmic singularity problem and is geodesically complete. This  bouncing scale factor can realize the canonical and two non-canonical scalar fields in the NOGHDE non-interacting viscous scenario (refer to Figs. \ref{Fig9_Vphi-Canonical}, \ref{Fig11_Vphi-Tachyon} and \ref{Fig13_Vphi-K-essence}). We also provide a solution which does not require inflation to the horizon problem, and show that the current patch comes into causal contact as the universe contracts before the bounce. To clarify the solution of the horizon problem by bouncing cosmologies and the model studied in this research, we shall explicitly compute the horizon size. As previously stated, the primary objective is to determine if the time derivative of the universe's total entropy, which is constrained by the horizons and the normal entropy, remains at a non-negative level. To do the same, we have imagined the universe to be encircled by a future event horizon. The event horizon has previously been included in order to determine the density of NOGHDE. The generalized second law is applicable to both pre- and post-bounce scenarios, regardless of the curvature of the universe, as we can observe from Figs. \ref{Fig15_Sdot}, \ref{Fig16_Sddot} when the model parameters are chosen appropriately although TE is established in post-bounce case only. 

In summary, bounce-inflation in terms of future event horizon can realize viscous non-interacting generalized hologrphic dark energy with Nojiri-Odintsov cutoff. In finalization, let us discuss the results of this study in the context of the recent work by \cite{new1}, whose authors proposed the holographic realisation of a cosmic scenario with an intermediate radiation-dominated era followed by a Kamionkowski-like reheating stage, from constant roll inflation (early on) to the dark energy era (later on). In light of \cite{new1}, we propose to extend our study to periods that genuinely follow the inflationary phase, demonstrating the unification of early inflation and late time acceleration.

\textcolor{black}{While concluding, let us comment on the relevance of the current work with respect to the existing literatures. Firstly, as holographic cut-off we adopted Nojiri-Odintsov cutoff, which is the most generalized cut-off. In this connection, this study is in line with the approaches adopted in \cite{Odintsov1,NO}. Furthermore, we have carried out the study in bulk viscosity framework following \cite{bulk2, bulk3}. We intend to further expand the outcomes of the work in modified gravity framework to investigate the possibilities of bounce inflation realisation with such a highly generalized infrared cut-off. }
\section{Acknowledgements}
\textcolor{black}{The authors sincerely thank the anonymous reviewer for the insightful comments.} Sanghati Saha is thankful to the Inter-University Center for Astronomy and Astrophysics (IUCAA), Pune, India, for providing all types of working facilities and hospitality, where most of this work was done from August 25 to September 6, 2023. Surajit Chattopadhyay acknowledges the visiting associateship of the Inter-University Center for Astronomy and Astrophysics (IUCAA), Pune, India.

\end{document}